\documentclass[twocolumn,english,prb,showpacs,preprintnumbers,amsmath,amssymb,aps]{revtex4}
\usepackage[T1]{fontenc}
\usepackage[latin9]{inputenc}
\usepackage{graphicx}

\makeatletter
\@ifundefined{textcolor}{}
{%
 \definecolor{BLACK}{gray}{0}
 \definecolor{WHITE}{gray}{1}
 \definecolor{RED}{rgb}{1,0,0}
 \definecolor{GREEN}{rgb}{0,1,0}
 \definecolor{BLUE}{rgb}{0,0,1}
 \definecolor{CYAN}{cmyk}{1,0,0,0}
 \definecolor{MAGENTA}{cmyk}{0,1,0,0}
 \definecolor{YELLOW}{cmyk}{0,0,1,0}
}

\makeatother

\usepackage{babel}
\begin{document}

\title{Bidirectional sorting of flocking particles in the presence of asymmetric
barriers}

\author{Jeffrey A. Drocco, C. J. Olson Reichhardt, and C. Reichhardt}

\affiliation{Center for Nonlinear Studies and Theoretical Division, Los Alamos
National Laboratory, Los Alamos, NM 87545}

\date{\today}
\begin{abstract}
We numerically demonstrate bidirectional sorting of 
flocking particles interacting with an array of asymmetric barriers.  
Each particle aligns with the average swimming direction
of its neighbors according to the Vicsek model and experiences additional 
steric interactions as well as repulsion from the 
fixed barriers. We show that particles preferentially
localize to one side of the barrier array over time, and that 
the direction of this rectification can be reversed by adjusting the 
particle-particle exclusion
radius or the noise term in the equations of motion. These results
provide a conceptual basis for isolation and sorting of single- and
multi-cellular organisms which move collectively according to flocking-type
interaction rules.
\end{abstract}
\maketitle

\vskip2pc

\section{Introduction}

The ensemble dynamics of self-driven particles 
can differ significantly from those
of Brownian random walkers \citealp{Hagen2011}. 
For example, in experiments on 
microfabricated habitats connected by funnel-shaped channels, 
self-propelled 
\textit{E. coli} bacteria
preferentially migrated to the chamber 
towards which the funnels pointed \citealp{Galajda2007,Galajda2008}
even though Brownian particles would have remained equally distributed in
both chambers.
A simple simulation model
showed that the rectification 
arises due to the modification of the run-and-tumble swimming dynamics
of the bacteria 
by the walls of the microenvironment
\citealp{Wan2008}. 
When 
a running bacterium encounters a wall, it does not reflect away from the 
wall
or immediately tumble, but 
swims in the direction of the wall while
preserving as much as possible its prior direction of motion. 
Refs. \citealp{Wan2008} and \citep{Tailleur2009} 
found rectification under this interaction 
rule for independent swimmers that did not interact with each other.
The rectification in the bacteria system
resembles a ratchet effect in which a net dc motion
occurs in the absence of a dc drive due to the application of an external
ac drive or flashing substrate \cite{reimann}. 
For self-driven particles, however, no external driving is necessary. 
In addition to demonstrations of directed bacterial motion achieved
through a ratchet mechanism \cite{kaehr09},
it has also been shown that baths of swimming bacteria can induce
directed rotational motion of
asymmetric flywheels
\cite{dileonardo,sokolov,angelani}.  

Interactions between self-propelled particles can lead to distinctive
dynamical behaviors that are more complex
than those of independently moving particles.
Simple models such as that of Vicsek {\it et al.} 
\cite{Vicsek1995,czirok97} 
capture many features of
the dynamics of species 
with strongly collective motion,
in which individuals preferentially align
with their neighbors and form moving groups
\citep{toner95,toner98}.
These models qualitatively reproduce the motion of both macro-scale
groups, such as fish schools and bird flocks \cite{Sumpter2006,Couzin2009}, 
and micro-scale
groups, such as bacterial swarms and cancerous tumors \citep{Deisboeck2009}.
The original Vicsek model
includes only a term for preferential velocity alignment with all neighbors
within a fixed flocking radius, yet it exhibits a phase
transition to unidirectional motion as a function of particle density
and noise amplitude.
Although numerous modifications of the Vicsek model have been proposed, 
such as the addition
steric interactions \cite{czirok96} and/or 
cohesion \cite{Couzin2002,gregoire04,orsogna06}, only a very limited
amount of work has been done on the interaction of flocking particles
with walls or barriers.  Walls can impose a directional symmetry
breaking \cite{toner98}, induce the formation of a vortex state
\cite{czirok96,duparcmeur95,szabo06,barbaro09}
or laning \cite{hernandezortiz09}, or simply serve as aggregation
focal points \cite{HO}; walls have also been used for
understanding
finite size effects \citep{Hoare2004,Hensor2005,hemelrijk10}, such as the
relationship between the collective dynamics of fish in a tank and
those of fish in the open ocean. 

In this work we simulate a modified version of
the Vicsek flocking algorithm that includes both steric repulsion between
particles and confinement within a two-dimensional microenvironment
with strategically placed gates similar to those of Ref.~\citep{Wan2008}.
Here we consider strictly repulsive particle-wall interactions, 
so that the particles
do not follow the walls when swimming independently.
As the particle density increases, we find rectification effects 
once the density is high enough to permit collective motion to
occur.
By varying the interparticle exclusion radius, the flocking radius, or the
noise, we can reverse the direction of the rectification. 
This result has implications for the potential
sorting of self-propelled particles that move according to these types
of interaction rules.

{\it Simulation--}
We consider a two-dimensional $L \times L$ system of $N$ self-driven particles 
at number density $\rho_0=N/L^2$
with fixed, repulsive boundaries on all sides. The 
overdamped equation of motion
for a single particle $i$ is
$d{\bf x}_i={\bf v}_{i}(t)dt$ with 
\begin{equation}
{\bf v}_{i}(t)={\bf f}_{vc}^{i}(t)+{\bf f}_{r}^{i}(t)+{\bf f}_{b}^{i}(t)
\end{equation}
All quantities are rescaled to dimensionless units. The time step
$dt=0.002$ and we take $L=66$. The velocity consensus force ${\bf f}_{vc}^{i}$,
also called the alignment force, 
is determined by the velocities
of all $M$ particles, including particle $i$, within a flocking
radius $r_{f}$ of particle $i$: 
\begin{equation}
{\bf f}_{vc}^{i}(t)=A_{f}\left(\cos(\Phi_{vc}^{i}(t)){\bf \hat{x}} +
\sin(\Phi_{vc}^{i}(t)){\bf \hat{y}}\right)
\end{equation}
with
\begin{equation}
\Phi_{vc}^{i}(t)=
\arctan^2\left(\sum_{j=1}^{M}\frac{{\bf v}_{j}(t-dt)}{|{\bf v}_{j}(t-dt)|}\right)+\xi .
\end{equation}
Here $A_f=2.0$ and 
$\xi$ is a random variable uniformly distributed on the interval
$[-\eta/2,\eta/2]$. Both the steric particle-particle interactions
${\bf f}_{r}^{i}$ and the particle-barrier interactions 
${\bf f}_{b}^{i}$ are given by the stiff spring repulsions:
${\bf f}_{r}^{i}(t)=\sum_{j \ne i}^{N}A_{r}(2r_{e}-r_{ij})\Theta(2r_{e}-r_{ij}){\bf {\hat{r}}}_{ij}$ 
and ${\bf f}_{b}^{i}(t)=\sum_{k}^{N_g}A_{p}(r_{e}+r_g-r_{ik})\Theta(r_{e}+r_g-r_{ik}){\bf {\hat{r}}}_{ik}$, 
where $A_{r}=200$, $A_p=10$, 
$r_{ij}=|{\bf r}_i(t)-{\bf r}_j(t)|$,
and ${\bf \hat{r}}_{ij}=[{\bf r}_i(t)-{\bf r}_j(t)]/r_{ij}$.
Here $r_{e}$ is the particle exclusion radius, $r_g=0.05$ is the barrier
exclusion radius, and there are $N_g=16$ barriers composed of the four
confining walls plus 12 V-shaped gates. 
${\bf r}_{ik}$ is the vector from the nearest
point on barrier $k$ to particle $i$, $r_{ik}=|{\bf r}_{ik}|$, and
${\bf \hat{r}}_{ik}={\bf r}_{ik}/r_{ik}$.
The length of each side of the V gates is
$L_B=4.9$ and 
the angle each V arm makes with the $y$ axis is 30$^\circ$. 
The spacing between the bases of the V's is 
$l_s=5.5$ 
and the spacing between the tips of adjacent V's is 
$l_o=0.6$. 
The 12 gates bisect the system into top and bottom chambers,
with the aperture of each funnel shape pointing toward the top chamber.
We initialize the system by distributing the particles at random throughout
the sample.
The equations of motion are then integrated for $3 \times 10^6$ 
simulation time steps.
In the absence of particle-particle interactions, the purely repulsive
wall interactions produce no rectification of the particles, in agreement
with the results of Ref.~\cite{Tailleur2009}

\begin{figure}
\includegraphics[width=3.4in]{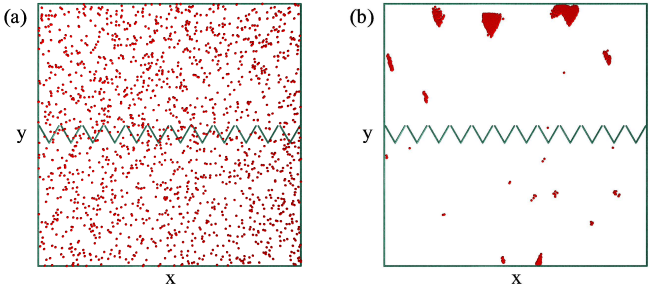} 

\caption{
Simulation images.  Lines: barriers and walls; dots: particle positions.
a) Initial state of sample with
$r_f=1.0$ and $r_{e}=0.07$ at $\rho_{0}=0.4$. 
b) The same sample
after $7 \times 10^5$ simulation time steps shows
rectification of particles into the top chamber.}

\label{fig:rectdemo} 
\end{figure}

{\it Results--}
In Fig.~\ref{fig:rectdemo}(a), we show an 
image of the simulation geometry in the randomly initialized state for
a system with $r_f=1.0$, $r_e=0.07$, and $\rho_0=0.4$.
After a sufficient amount of time elapses, the particles 
concentrate in one of the two chambers, reaching a steady state value 
of $\rho_{top}$, the density in the top chamber.
In Fig.~\ref{fig:rectdemo}(b), after $7 \times 10^5$ simulation time steps
the particle density is clearly higher in the top chamber.

\begin{figure}
\includegraphics[width=3.5in]{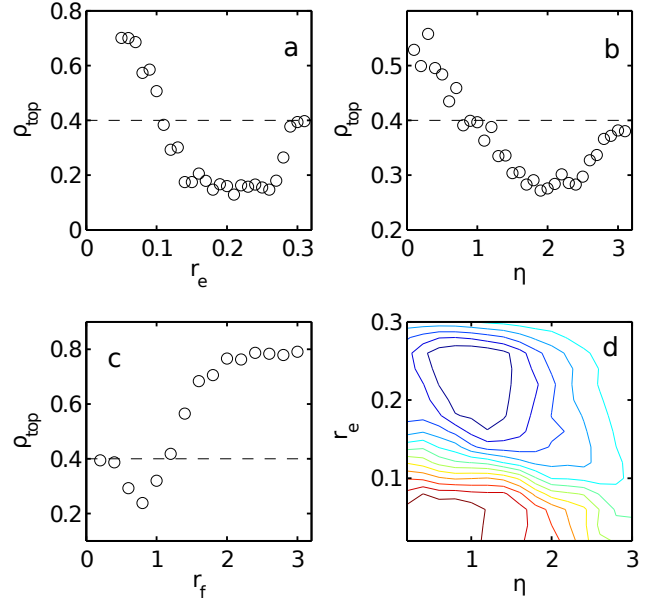} 

\caption{
(a-c) $\rho_{top}$, the density in the top chamber, after $3 \times 10^6$ 
simulation time steps for a sample with initial density $\rho_{0}=0.4$, 
indicated by the dashed line.
(a) $\rho_{top}$ vs $r_e$ for $\eta=1.5$ and $r_f=1.0$.
The rectification reverses at $r_e=0.12$ and drops to zero for $r_e\geq 0.3$.
b) $\rho_{top}$ vs $\eta$ for $r_e=0.12$ and $r_f=1.0$.
The rectification reverses at $\eta \sim 1.0$ and drops to zero for
$\eta \gtrsim \pi$.
c) $\rho_{top}$ vs $r_f$ for $r_e=0.12$ and $\eta=1.5$.
For $r_f<r_e$ only steric particle interactions occur and rectification is
negligible.  There is a rectification reversal at $r_f \approx 1.2$, and
for large $r_f$ when all the particles tend to align into a giant flock,
the particles accumulate in the top chamber.
d) Rectification phase diagram for $r_e$ vs $\eta$.  Lower contours (red)
indicate rectification into the top chamber and upper contours (blue)
indicate rectification to the bottom chamber.
}

\label{fig:flocknoiserad} 
\end{figure}

We find that we
can vary whether the rectification moves the particles into the
top ($\rho_{top}>\rho_0$) or bottom ($\rho_{top}<\rho_0$) chamber by altering
$r_{f}$, $r_{e}$, or $\eta$, as shown in
Fig.~\ref{fig:flocknoiserad}(a-c) where we plot $\rho_{top}$ after
$3 \times 10^6$ simulation time steps. 
For small values of $r_{e}$ and $\eta$, particles are
rectified into the top chamber, but a rectification reversal
occurs at $r_e=0.12$ and $\eta\sim 1.0$ in Fig.~\ref{fig:flocknoiserad}(a)
and (b), respectively. 
There is a saturation into a nonrectifying state
for $r_e\ge 0.3$ in Fig.~\ref{fig:flocknoiserad}(a); this corresponds
to $2r_e \ge l_o$ and occurs when the particle diameter becomes larger
than the aperture between adjacent gates, so that particles can no longer
pass between the upper and lower chambers.
In Fig.~\ref{fig:flocknoiserad}(b) rectification vanishes for 
$\eta \gtrsim \pi$ when
the alignment force between neighboring
particles, and thus the tendency of particles to form flocks, 
is almost completely destroyed by noise.
We plot a rectification phase diagram 
as a function of $r_{e}$ and $\eta$ in Fig.~\ref{fig:flocknoiserad}(d),
showing that rectification into the upper chamber occurs for small values
of $r_e$ and $\eta$, while reversed rectification into the lower chamber
appears for larger $r_e$ and small $\eta$.

We next consider the flocking radius $r_f$.  Fig.~\ref{fig:flocknoiserad}(c)
indicates that no rectification occurs when $r_f<r_e$.  In this limit,
the particles interact only sterically and have no flocking interaction,
and the repulsive barrier walls produce no rectification in the absence
of flocking.  For $r_e<r_f<1.2$, we find a reversed rectification into the
lower chamber, while for all $r_f \geq 1.2$, the particles rectify into the
top chamber.  For $r_f>2.0$, the value of $\rho_{top}$ saturates at
$\rho_{top}=0.8=2\rho_0$, indicating that nearly all of the particles are
located in the top chamber.

\begin{figure}
\includegraphics[width=3.5in]{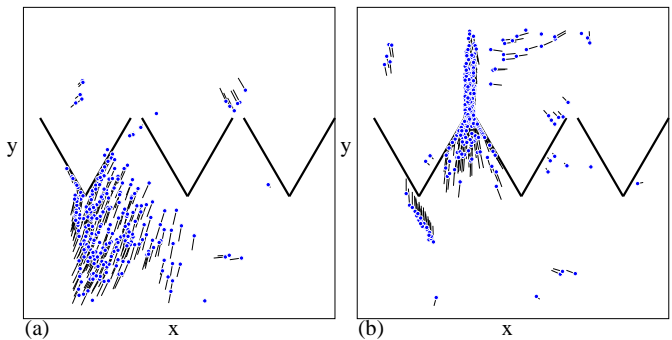} 

\caption{
Illustration of rectification into the top chamber
for low noise $\eta=1.5$ and small exclusion radius $r_e=0.05$
at $r_f=1.0$.  
A $15 \times 15$ section of the sample is shown.  Dots: particle
positions; light lines: particle trajectories; heavy lines: barriers.
A flock incident on the gates from the
bottom chamber (a) condenses and elongates in order to file through the
aperture between gates (b). Flocks incident on the gates from the top
chamber 
have a much lower probability of passing through the aperture and
cannot be funneled into a similar oblong shape.}

\label{fig:forwarddiagram} 
\end{figure}

\begin{figure}
\includegraphics[width=3.5in]{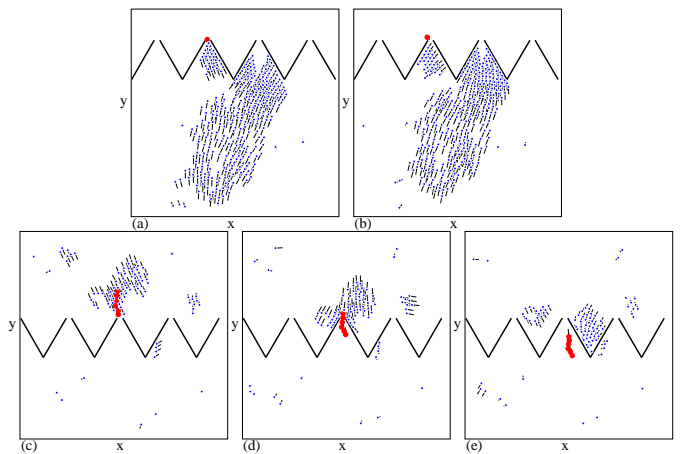} 

\caption{
Illustration of rectification into the lower chamber
for low noise $\eta=1.5$ and large exclusion radius $r_e=0.2$
at $r_f=1.0$.  A $20 \times 20$ section of the sample is shown.  Dots:
particle positions; light lines: particle trajectories; heavy lines:
barriers.
(a,b) Flocks incident on gates from the bottom chamber cannot fit through
the aperture for this value of $r_e$; the flock jams inside the funnel
while a single particle (highlighted in red) escapes from the flock and
enters the upper chamber.
The remainder of the flock returns to the lower chamber.
(c,d,e) Flocks incident on gates from the top chamber are fragmented
and a small group of particles (highlighted in red) can escape from 
the flock and enter the lower chamber.
The flock fragmentation process occurs with greater frequency as the 
flocks become less cohesive due to either higher $\eta$ or higher
ratios $r_e/r_f$.
}

\label{fig:reversediagram} 
\end{figure}

The rectification reversal occurs due to a 
change in the nature of the microscopic
interaction between the flocks and the funnel channels.
For example, as the exclusion radius $r_{e}$
increases, the particles are less able 
to form tight and cohesive flocks. At
low values of $r_e$, particles are rectified into the top
chamber when
flocks, incident on the gates from the bottom, rearrange
into oblong shapes and pass efficiently
through the funnel, as illustrated in
Fig. \ref{fig:forwarddiagram}.
For higher values of $r_e$, the steric interparticle repulsion 
prevents the flocks from condensing and makes it impossible for more
than one particle at a time to pass through the funnel aperture.
As a result, the particles clog inside the funnel rather than passing
through, as illustrated in Fig.~\ref{fig:reversediagram}(a).
The flock reverses direction due to the repulsion from the barrier
walls, and at most one or two particles occasionally manage to escape the flock
and enter the top chamber, as shown in 
Fig.~\ref{fig:reversediagram}(b).
In contrast, a flock that approaches the gates from the upper chamber is
fragmented by the gates into two smaller flocks; when this occurs,
particles that are directly incident on the aperture between gates can
escape from both flocks and pass in a single file into the lower
chamber, as illustrated in Fig.~\ref{fig:reversediagram}(c,d,e).
Since the average number of particles escaping the flock and crossing the
barrier is larger when the flocks are approaching from above than when
they are approaching from below, a net rectification into the lower
chamber occurs over time.
We note that the reversed rectification into the lower chamber 
(Fig.~\ref{fig:reversediagram}) is a much
slower process than the forward rectification into the higher chamber
(Fig.~\ref{fig:forwarddiagram}), although we are able to reach a
steady state within our simulation time for either process.
In spite of this, we 
find that the maximum possible amount of rectification that can
be achieved in steady state as the parameters are varied
is the same for both directions of rectification, as shown in
Fig. \ref{fig:flocknoiserad}(a).

\begin{figure}
\includegraphics[width=3.5in]{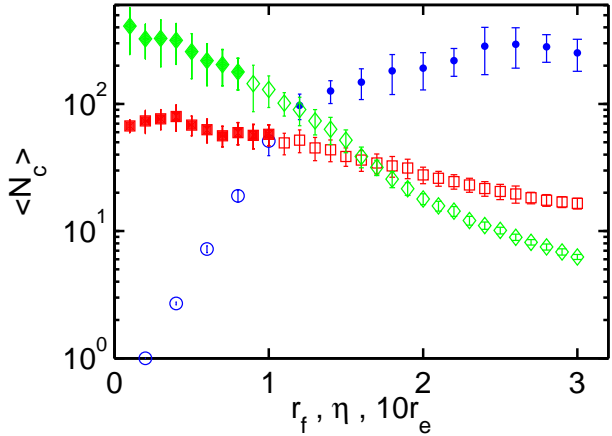} 

\caption{
Mean flock size $N_c$, in number of particles, 
vs $r_{f}$ (blue circles), $r_{e}$ (red squares), and $\eta$ (green diamonds). 
Error
bars indicate standard deviation. 
}

\label{fig:clustering} 
\end{figure}

Reversed rectification into the lower chamber also occurs whenever the
flocks become fragile or prone to breakage.  This occurs both when the
noise parameter $\eta$ is increased and when the flocking radius
$r_f$ is reduced.  Under these conditions, the flocks are not
cohesive enough to flow as a unit through the funnel aperture in the
manner illustrated in Fig.~\ref{fig:forwarddiagram}; at the same time,
the probability that a flock will fragment and lose some of its
members to the lower chamber when approaching the gates from above, as
in Fig.~\ref{fig:reversediagram}(c-e), is increased.
In Fig.~\ref{fig:clustering}(a), we plot the average flock size $N_c$ as
a function of $r_f$, $r_e$, and $\eta$ for the systems in 
Fig.~\ref{fig:flocknoiserad}.  
We separate the particles
into clusters iteratively by identifying particles that are within the
flocking radius $r_f$ of each other;
$N_c$ is then the average number of particles per
cluster.  The value of $N_c$ is higher in regimes where the particles are
rectified to the top of the container, and lower in the reversed rectification
regime.

\begin{figure}
\includegraphics[width=3.4in]{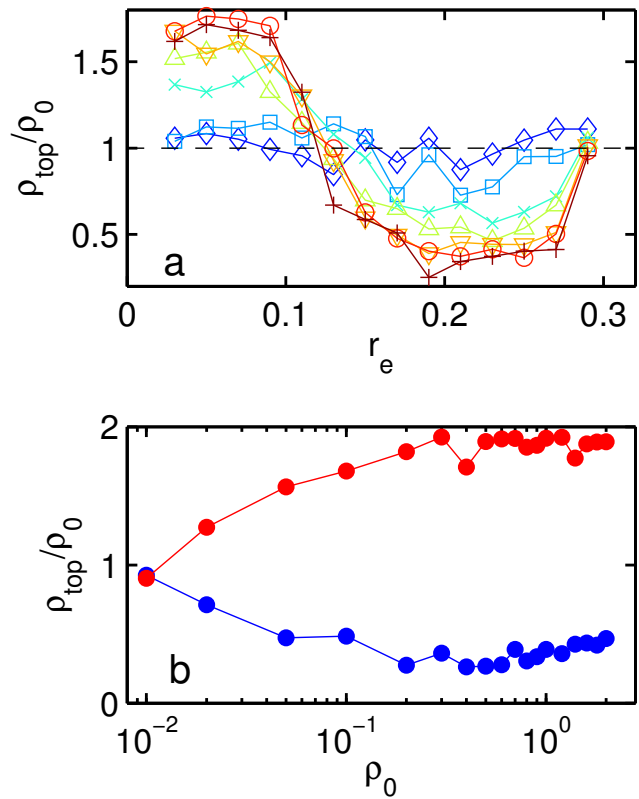} 

\caption{
Dependence of rectification on initial particle density $\rho_0$
for a system with $r_f=1.0$ and $\eta=1.1$.
(a) 
$\rho_{top}/\rho_0$
after $3 \times 10^6$ simulation time steps vs $r_e$ for different values
of $\rho_0$.  From blue to red, 
$\rho_{0}=0.004$ ($\diamond$), 
$0.01$ ($\Box$), 
$0.03$ (x), 
$0.05$ ($\bigtriangleup$), 
$0.08$ ($\bigtriangledown$), $0.1$ ($\bigcirc$), $0.12$ (+).
A rectification reversal 
emerges as $\rho_{0}$ increases. 
b) 
$\rho_{top}/\rho_0$ after $3\times 10^6$ time steps vs $\rho_0$ for
$r_{e}=0.05$ (upper red curve) and $r_{e}=0.23$ (lower blue curve).}

\label{fig:densplotter} 
\end{figure}

One of the unique aspects of the rectification behavior described
here is that, unlike previous rectification phenomena reported
for self-driven particles \citealp{Galajda2007}, it
occurs only when the 
initial particle density $\rho_0$ is high enough for flock formation
to occur.
In the limit of low $\rho_0$,
when the particles are moving independently and
not able to form flocks, individual particles
simply reflect off the barriers
in a manner similar to inertial particles. 
This type of barrier interaction has been shown to produce
no rectification in 
noninteracting particle limit \citep{Tailleur2009}, and 
as indicated in Fig.~\ref{fig:densplotter}(a) we find
no rectification at low densities $\rho_0<0.01$.
As $\rho_0$ increases, both rectification and a rectification reversal
emerge, and the amount of rectification saturates for
$\rho_{0}\geq 0.1$, as shown in
Fig.~\ref{fig:densplotter}(b). 
We note that since $\rho_{0}$ represents
number density, rather than surface area covered, 
it is possible to have $\rho_0>1$.

{\it Conclusion--}
We have implemented a simple model of flocking particles in the presence
of fixed, repulsive barriers, and find that such particles will concentrate
on one side of a set of asymmetric V-shaped gates. The direction
of the rectification can be reversed by modulating any of three parameters:
the flocking radius $r_{f}$, the exclusion radius $r_{e}$, or the
noise parameter $\eta$. The existence of the rectification and its
direction are determined by the ability of the particles to form
flocks and the robustness of the flocks against breakage; in the low
density limit, when no flocks appear, we find no rectification due to the
purely repulsive interactions of the particles with the barrier walls.
Thus, the rectification we observe arises strictly due to collective
effects.
The bi-directional rectification behavior we describe could be used
to sort particles which tend to concentrate on different sides of
the barrier \cite{epaps}.
This
effect is similar to the sorting phenomenon observed by Mahmud \textit{et
al.} for cancer cells\citep{Mahmud2009}. We expect sorting devices
based on these principles to have broad potential applications with
regard to both biomedical diagnostics and therapeutics.

This work was carried out under the auspices of the 
NNSA of the 
U.S. DoE
at 
LANL
under Contract No.
DE-AC52-06NA25396.

\begin{figure}
\includegraphics[width=3.4in]{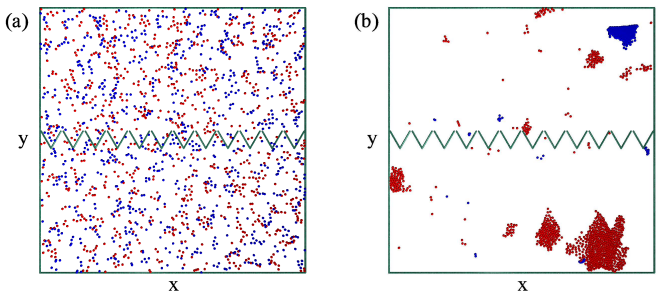}
\caption{Supplemental Figure.  Demonstration of two-species sorting. 
Simulation of system containing 871 "A" particles with $r_{e}=0.22$ (green) and 871 "B" particles with $r_{e}=0.055$ (pink). $r_{f}=1.0$ and $\eta=1.1$ in both cases.  Particles of different species repel via steric repulsion but only experience alignment forces with particles of the same species. Simulation shown (a) at time $t=0$ and (b) after $4 \times 10^6$ simulation time steps when many of the "A" particles have rectified into the bottom chamber and most of the "B" particles have rectified into the top chamber. The bidisperse system requires longer times to reach a steady state compared to the monodisperse system.}
\end{figure}

\end{document}